# Assessing electron emission induced by pulsed ion beams: a time-of-flight approach

S. Lohmann [a], A. Niggas [b], V. Charnay [a], R. Holeňák [a], D. Primetzhofer [a]

[a] *Department of Physics and Astronomy, Uppsala University, Box 516, 751 20 Uppsala, Sweden*

[b] *Institute of Applied Physics, TU Wien, 1040 Vienna, Austria*

**Abstract**

We present a method to measure the kinetic energy of electrons emitted upon ion impact via their time-of-flight. Pulsed beams of $H^+$, $D_2^+$, $He^+$ and $Ne^+$ ions with velocities between 0.4 and 3.5 a.u. are transmitted through thin, self-supporting carbon and gold foils. Transmitted ions and secondary electrons are detected with a position-sensitive detector behind the sample and their respective energies are determined via their flight times. A coincidence criterium can be applied in the acquisition software. Measured electron energies range between 10 and 400 eV. Above ion velocities of 1 a.u. the most probable electron energy scales with ion velocity pointing towards a kinetic emission mechanism. At lower ion velocities, the electron energy stays constant and lies above the maximum energy transfer possible in a classical binary collision between ion and electron. Potential applications and technical challenges of measuring electron energies and yields with a time-of-flight approach are discussed.

## 1. Introduction

In the interaction of ions with matter electron emission is one of the most fundamental processes observable, and it has therefore been studied for decades. Understanding characteristics such as electron energy distributions, electron yields and electron emission statistics is essential for predicting plasma-surface interactions in fusion reactors [1,2], imaging with the helium ion microscope [3,4] or efficient signal amplification [5]. With growing interest in 2D materials such as graphene, the role of ion-induced electron emission as a probe for the electronic response can be of relevance for the design of new electronic devices [6,7].

Ion-induced electron emission is often categorised according to the energy transfer mechanism. Kinetic emission means that kinetic energy from the incident ion is transferred to an electron in a binary collision [8]. Potential emission, on the other hand, is an umbrella term for a number of processes, in which potential energy stored in the ion is converted into an electronic excitation of the target [9,10]. Irrespective of their specific origin, electrons with enough kinetic energy can produce tertiary electrons in collision cascades. Whereas kinetic emission is the dominating process for fast ions ($v > 1$ atomic unit (a.u.)), the use of slow and/or highly charged ions leads to primarily potential emission. By studying the electron emission characteristics as a function of incident ion velocity, information on the specific nature of the energy transfer mechanism between ion and the electronic system of the target can be obtained.

Electrons emitted upon ion impact are usually detected with the help of magnetic or most commonly electrostatic spectrometers [11]. For accurately assessing even small electron yields, the electron emission statistics is measured by subjecting the electrons to electric potentials of several kV [12]. A set of retarding grids can be used to analyse the energy distribution in this case [13], but in general the emission geometry and information on electron energy are not easily accessible with this approach. In principle, the electron energy can be measured via the time-of-flight (ToF) method if pulsed ion beams are employed. To our knowledge, however, very little effort has been made into this direction and existing experiments still use guiding by magnetic fields, which makes the determination of flight time and emission direction much less direct [14,15].

We present a method to detect secondary electrons in coincidence with primary ions transmitted through thin, self-supporting foils. The energy of electrons as well as transmitted ions is determined via their respective ToF. Additional position information is available by the use of a large solid-angle delay-line detector. We give a detailed description of our set-up and experimental approach in Section 2 and present first results in Section 3.

## 2. Experimental methods and data analysis

Experiments were performed at the time-of-flight medium energy ion scattering set-up at Uppsala university [16,17]. Beams of $H^+$, $D_2^+$, $He^+$ and $Ne^+$ ions were provided by a 350 kV implanter and electrostatically chopped to pulses with a chopping frequency of 1/32 MHz. Pulse widths are 1 ns to 2 ns and the beam current impinging on the sample is 2-3 fA. A sketch of the relevant parts inside the experimental chamber is given in Fig. 1. The pressure inside the vacuum chamber is on the order of $1 \cdot 10^{-8}$ mbar. Note that the chamber is not shielded against external magnetic fields. Potential influences of the earth magnetic field are discussed later in the text.

We employed a position-sensitive microchannel plate (MCP) detector to simultaneously detect photons, electrons and ions. The detector can be rotated freely around the centre of the scattering chamber (40 mm from sample position), thus, enabling experiments in different scattering geometries. For the results presented in this work, ions were transmitted through thin, self-supporting foils and detected together with ion-induced secondary electrons 250 mm behind the target. The detector covers a solid angle of 0.13 sr, and the position of particles is determined with help of two delay lines orthogonal to the primary beam direction. The energy of ions as well as electrons is measured via their respective flight time. In order to efficiently detect even low-energy electrons, the front MCP was kept at a potential of $V_{MCP}$ = +200 V. 18 mm in front of the MCPs a nickel grid (90 % transmittance) is mounted, whose potential $V_{grid}$ can be adjusted. If not stated otherwise, however, $V_{grid}$ = 0 V for all presented experiments. To accelerate electrons towards the detector, a bias voltage can also be applied to the sample holder. The influence of this potential $V_{sample}$ on the electron spectrum will be discussed later in the text. A shielding cone was added to the sample holder to achieve a better uniformity of the potential at the point, where electrons leave the sample.

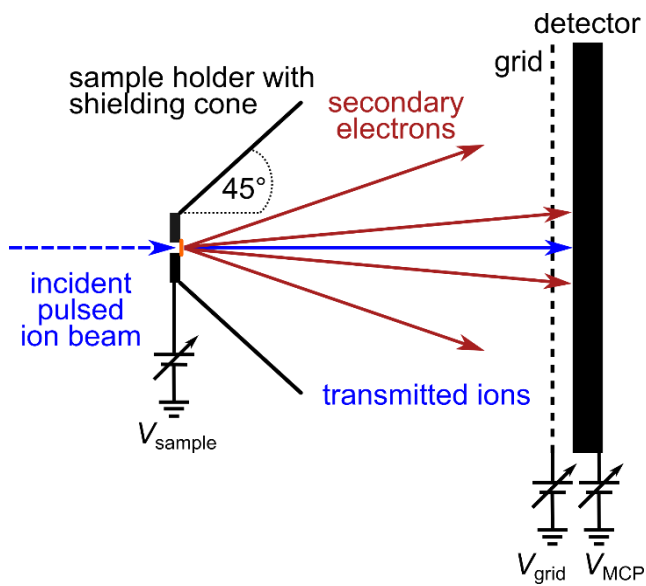

Figure 1: Schematic cut along the beam axis visualising the relevant geometries and potentials inside the scattering chamber. The detector consists of two stacked MCPs and two delay lines, and simultaneously records the flight time and two position signals of transmitted ions, secondary electrons and photons. $V_{MCP}$ here refers to the potential of the front MCP, which was kept at +200 V for all presented experiments. The other two potentials were varied. Note that the sample holder and the shielding cone lie on the same potential $V_{sample}$. Not to scale, see text for information on the actual distances.

We used gold foils with a thickness of 141 nm and carbon foils of three different thicknesses (25 nm, 40 nm and 50 nm) as samples. The Au foil areal density was determined with Rutherford backscattering spectrometry, and the C foil thicknesses are given as specified by the manufacturer (Micromatter Technologies Inc.). A more precise knowledge of the sample areal densities is not relevant for this study due to the measurement method (see Section 3). Samples were floated on water and then transferred to the sample holder, where they keep sticking upon drying. The respective foil covers an opening in the holder of 2.5 mm diameter and thereby forms a self-supporting target. No sample cleaning was performed prior to experiments. We are aware of the influence of surface conditions on the emitted secondary electron spectrum [8], but we aim to present a measurement concept rather than analysis of a specific material in this work. Nevertheless, sample surfaces were characterised with Auger electron spectroscopy. The surface of the C foils contained between 6 and 12 % oxygen, and the gold surface was found to be covered by around 83 % C and 3.4 % O. We, therefore, expect similar results from all samples for a surface sensitive mechanism such as electron emission, but the higher stopping power and thickness of the Au foil allowed us to probe different ion velocities.

Figure 2a shows a time-of-flight spectrum recorded for a 100 keV $D_2^+$ beam and a C target. No voltage was applied to the sample holder. Prompt photons that are emitted from the target upon ion impact are detected at 0.8 ns, and serve for time calibration [18]. Between flight times of about 25 and 110 ns a broad distribution of alleged electrons is visible. They arrive before D ions transmitted through pinholes in the sample (peak at 116 ns) and the C sample itself (peak at 121 ns). Note that this electron distribution is only detected with a positively biased front MCP. For a standard MCP detector configuration with $V_{MCP}$ < -2 kV only photons and ions/neutrals with keV energies are detected (see e.g. [16] for such a ToF spectrum). In Fig. 2b a time-to-energy converted electron spectrum is depicted (corresponds to the shaded area in 2a). The full and the dashed vertical line indicates the position of the most probable and the mean electron energy, respectively. We also calculated the maximum energy transfer $T_{max}$ from an ion (energy $E_{ion}$, mass $m_{ion}$) to an electron (mass $m_e$) in a classical binary collision with recoil angle φ according to:

$$T_{\max}(\varphi = 0) \approx \frac{4m_e}{m_{\text{ion}}} E_{\text{ion}}. \tag{1}$$

The result for 45.9 keV $D^+$ is given by the dotted line. For a full discussion of these various energy notions see Section 3. The inset in Fig. 2b shows the spatial distribution of electrons on the detector.

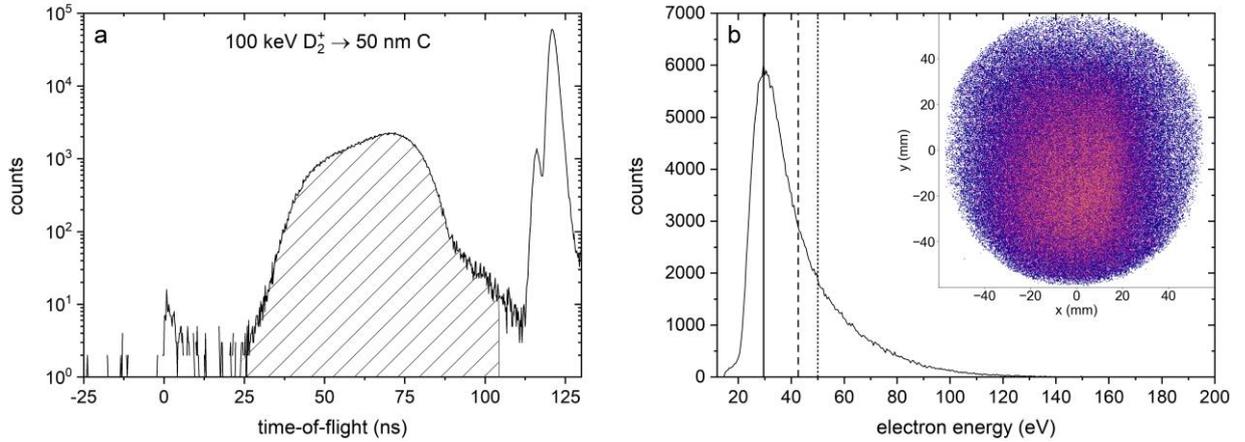

Figure 2: a: Time-of-flight spectrum measured using 100 keV $D_2^+$ ions as projectiles and a 50 nm self-supporting C foil as the sample. A prompt peak of photons, a broad distribution of secondary electrons (shaded area) and primary ions transmitted through pinholes (peak at 116 ns) and the sample itself (peak at 121 ns) are visible. b: Time-to-energy converted electron spectrum. The position of the most probable and the mean electron energy is visualised by the full and the dashed vertical line, respectively. The calculated maximum energy transfer from 45.9 keV $D^+$ to an electron in a classical binary head-on collision is indicated by the dotted line. The inset depicts the spatial distribution of electrons on the detector.

An energy spectrum as shown in Fig. 2b is derived by assuming a straight electron trajectory from the sample to the recorded position on the detector. We further assume that the electrons drift with constant velocity towards the grid, and are subsequently linearly accelerated between the grid and the front MCP. For ion velocities above ~1.5 a.u. the position detection becomes faulty since the run time of the delay line is on the order of the time difference between the arrival of an electron and the associated ion. For these cases, we therefore evaluated only the timing signal from the MCP assuming a mean flight trajectory length of 250 mm.

On the software side we can also choose to consider only electrons that were detected in coincidence with an ion transmitted through the sample. For slow ions, adding such a coincidence criterium does not change the shape of the energy spectrum and only reduces the intensity by a few percent. For faster ions, however, a reduced detection efficiency for signals arriving shortly after each other would greatly reduce statistics and increase measurement time and, therefore, no coincidence criterium was applied. From the results obtained at lower ion energies we are, however, confident of the electrons being ejected as secondary products of the ion-sample interaction.

Electron emission can also be studied in backscattering geometry with our set-up. However, due to the low backscattering probability for ions, detected secondary electrons cannot be correlated with an ion. In addition, the azimuthal symmetry of the experiment is lost in this geometry.

## 3. Results and discussion

To further confirm that the broad distribution observed in Fig. 2b are electrons emitted from the sample upon ion impact, we have studied the influence of the sample voltage on the ToF spectrum. Figure 3 shows spectra recorded by employing 200 keV He ions as projectiles and the Au foil as the sample for different voltages applied to the sample holder. The intensity of all spectra is normalised to the peak of transmitted ions (at 103 ns flight time). Whereas this peak of keV ions is not detectably influenced by the applied voltages, the alleged electron distribution is indeed shifted towards shorter flight times for more negative $V_{sample}$. The peak shape also becomes narrower in the time domain as expected for adding a constant energy to a given initial distribution. The observed dependency of the flight time on the accelerating voltage behaves qualitatively similar as previously studied for positive desorbed ions with the same set-up. The model applied in [19] and [20] assumes initial acceleration close to the sample followed by a long drift towards the grid and final acceleration between grid and MCP. The initial energy of ions can then be fitted as a free parameter and was found to be of the order of a few eV only. The data shown in Fig. 3 can be fitted reasonably well with this model assuming electrons with initial energies close to the most probable energy obtained from a time-to-energy converted spectrum measured without an applied field (as shown in Fig. 2b). Additionally, we have performed simulations of electron trajectories with the SIMION software package [21]. By using measured most probable electron energies as initial input parameters, we obtained flight times and spatial distributions on the detector similar to our experimental results. We, therefore, conclude that we indeed observe ion-induced electron emission from the sample.

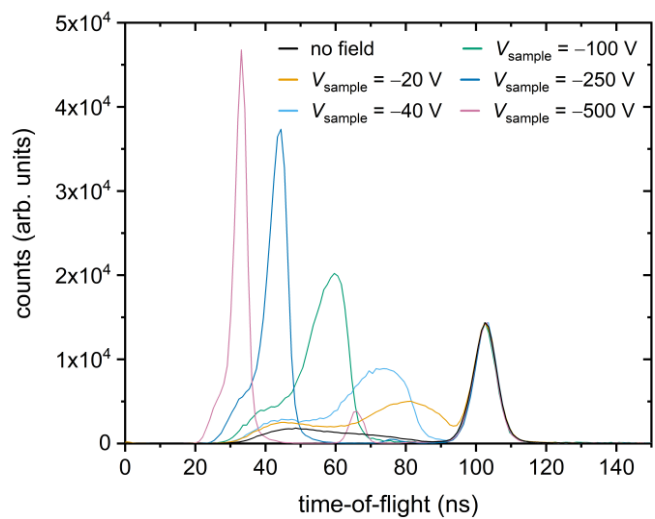

Figure 3: Influence of different negative sample bias on the secondary electron ToF spectrum. All spectra are recorded by transmitting He$^+$ with initial energy of 200 keV through a 141 nm thick Au foil, and all intensities are normalised to the respective transmitted ion peak. The grid was kept grounded for all shown measurements.

To study the ion-electron interaction in more detail and potentially learn details about the emission process, we measured the electron energy for a wide range of different ion species and ion velocities as well as for different sample materials. All results are shown in Fig. 4. Since electrons are detected in transmission geometry and expected to originate from the near-surface layers at the backside of the sample, electron energies are given as a function of the ion velocity after transmission. The exit ion energy is, hereby, directly determined from the flight time and, therefore, no information on the actual sample areal density or electronic stopping power is needed.

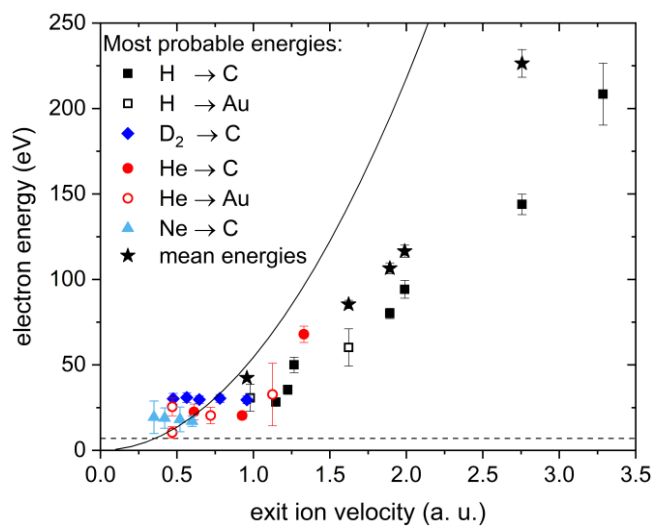

Figure 4: Electron energies measured with a ToF approach as a function of transmitted ion velocity. Shown are most probable electron energies for protons (black squares), D$_2^+$ (blue diamonds), He$^+$ (red circles) and Ne$^+$ ion projectiles (light blue triangles) using self-supporting C (closed symbols) and Au samples (open symbols). Black asterisks give the mean electron energy for selected measurements. The full line represents the maximum energy transfer from the ion to an electron in a binary head-on collision. The horizontal dashed line indicates the minimum electron energy necessary to reach the detector despite deflection by the earth magnetic field.

All data points except for black asterisks give the most probable electron energy (conf. full vertical line in Fig. 2b). We present results for protons (black squares), $D_2^+$ (blue diamonds), $He^+$ (red circles) and $Ne^+$ ions (light blue triangles). All full symbols correspond to C samples, whereas open symbols represent data obtained using the Au foil. The mean electron energy is indicated by black asterisks for selected measurements. The full black line gives the maximum energy transfer $T_{max}$ calculated from Eq. (1).

The error bars include the statistical uncertainties of the flight time, based on the time resolution of the system, and the sample-detector distance due to positioning of the sample and the finite size of the beam spot. Additionally, a systematic error due to the influence of the earth magnetic field is expected, because the experimental chamber is not shielded against external magnetic fields and the energy of detected electrons is low. Assuming a local magnetic field strength of 51 000 nT and an inclination of 73° [22], the gyroradius of an electron with 10 eV (200 eV) kinetic energy is for example 210 mm (950 mm), and electrons are primarily deflected towards the right. Since the sample-detector distance is 250 mm, we must, therefore, conclude that the detected electrons, depending on their energy will follow trajectories with different curvature. The consecutive difference in trajectory length between a straight and a bent flight path introduces a systematic, energy-dependent error to the energy determination, which is also included in the error bars. The exact influence of the magnetic deflection depends on the emission angle as well but considering the small detection angle, we expect this additional systematic error to be minor for all but the very lowest detected energies. The deflection of electrons in the earth magnetic field also implicates that particles with a too small gyroradius will not reach the detector at all. For our geometry this minimum gyroradius at an emission angle of ~45° is 181 mm, which corresponds to an electron energy of 7 eV. Electrons with lower energies cannot be reasonably detected with our current approach and set-up even under optimum emission angle. This lower detection limit is also drawn as a horizontal dashed line into Fig. 4. As a consequence of this argument the second apparent electron peak appearing in Fig. 3 for higher applied sample bias is unlikely to originate from the sample and is, therefore, not further considered in this work.

In all cases the electron energy spectrum looks similar to the one plotted in Fig. 2b. The electron distribution is broad and features a long high-energy tail. Noteworthy, we have not detected a peak at the position of the C Auger line (272 eV [23]) for any of the employed probe beams. No other distinct features apart from the maximum in intensity are observed within the experimental statistics, either. The energy distribution of electrons emitted from the sample depends on a multitude of factors and disentangling them from each other will not directly be possible with our experimental method. First, the ion can excite electron-hole pairs at different impact parameters resulting in electrons with different energies and directions. Collective excitations can, in addition, generate electrons with higher energies than classically possible. Electrons can scatter again, thereby, losing energy and changing direction and form collision cascades, which leads to an exponential-like behaviour at the high-energy side of the peak. Second, only electrons with sufficient energy can actually leave the sample and third, without any directional electric field only electrons ejected into the solid angle covered by the detector contribute to the spectrum shown in Fig. 2b. Noteworthy, for ions in the studied velocity regime, momentum transfers in forward direction will always lead to final electron momenta outside the initial Fermi sphere, i.e. electron emission [24]. At the low-energy side the spectrum is additionally cut off due to the influence of the earth magnetic field as discussed above. For all these reasons the most probable electron energy cannot be assigned to a clear physical variable. It is merely the most likely outcome combining the above-mentioned processes.

At ion velocities above ~1 a.u. the electron energy increases for increasing ion velocity. The most probable energies lie well below the curve given by Eq. (1) due to the energy attenuation explained above. Mean electron energies are significantly higher than most probable energies but follow a similar velocity scaling. This behaviour supports our choice to evaluate the most probable energy since it scales with the ion velocity in a similar way than the whole energy distribution. At low exit ion velocities (1 a.u. and below) the most probable electron energy no longer scales with velocity, but is constant within the measurement uncertainty. We observe a small dependence on the projectile type – the electron energy ranges from about 18.5 eV for neon to about 30 eV for deuterium projectiles. Note that we do not detect an isotope effect for H and D at 1 a.u. where the respective ranges of studied ion velocities overlap. In both velocity regimes no material dependence is observed.

The velocity scaling at higher ion velocities strongly indicates that the majority of primary excitations is due to electrons excited in binary collisions with the primary ions. However, since electrons in the observed energy range have an inelastic mean free path of about 10 Å only [25], most secondary electrons will scatter and produce tertiary electrons with lower energy. Since kinetic emission in head-on collisions is the most directional of the present mechanisms and our experimental approach detects electrons in forward direction, a signature of this energy transfer should remain in the energy distribution, which we indeed observe even though the majority of electrons leaving the sample will be originated from these electronically driven cascades. A broad energy distribution as plotted in Fig. 2b, which peaks at an energy well below that of an electron directly ejected into the vacuum by a direct head-on collision, is the result. To support this argument, we have additionally measured the electron energy distribution for different detector angles in forward direction. No pronounced emission-angle dependence could be observed, which points again to tertiary electrons and not electrons ejected via direct close collision, whose energy depends on the recoil angle.

The short inelastic mean free path also means that we primarily detect electrons generated close to the surface and do not expect any influence of the sample bulk. According to the Auger electron spectroscopy results surfaces of all samples consist mostly of (hydro-)carbon (compare also with mass spectra presented in [20]). Therefore, we see no significant difference between the C and Au foils.

For the lowest studied ion velocities, the most probable electron energy is higher than the maximum energy transfer possible in a binary collision as given by Eq. (1). Therefore, an additional electron ejection mechanism different from kinetic emission needs to be considered. A possible explanation could be a contribution from electrons emitted via plasmon decay. Results from electron energy loss measurements on graphite show indeed a plasmon involving all valence electrons at an energy of 27 eV [26], which coincides well with the electron energies measured at lowest ion velocities.

In principle our current approach allows to measure electron yields together with the electron energy, however, these measurements are impaired by two technical difficulties. First, the already mentioned influence by the earth magnetic fields implies that with the current set-up not all electrons can be detected. This problem can be at least partially circumvented by applying a sufficiently strong sample bias. Detected yields indeed increase from around 0.25 electrons per incident ion without applied field to for example 1.4 (for 50 keV H$^+$) or even 2 (for 50 keV D$_2^+$) for $V_{sample}$ = -350 V. Note that reported yields for H transmitted through C foils are on the order of 3-4 in this energy regime [27]. Second, our present detection system has a dead time between 10-20 ns even if no position signal is acquired [28]. Therefore, yield measurements will be inaccurate if multiple electrons with similar energy are emitted per ion.

## 4. Summary and outlook

We have demonstrated a direct approach to measure the energy of electrons emitted upon the impact of keV ions via their flight time. We presented results from transmission experiment using $H^+$, $D_2^+$, $He^+$ and $Ne^+$ ions and self-supporting C and Au foils, and the energies of detected electrons lie between 10 and 400 eV. Above exit ion velocities of about 1 a.u. the detected electron energy scales with ion velocity, which points towards kinetic emission, whereas the constant electron energy at low ion velocities may indicate that additional different emission mechanisms need to be considered.

A target preparation chamber attached to the current experimental chamber is currently under construction. In the future, experiments on better defined surfaces will therefore be possible. By using single-crystalline, self-supporting samples, electron emission along channelling and random trajectories could be compared. Even the response of 2D materials to light ions could be measured. In this scenario, a possibility would be to detect electrons in coincidence with ions at large scattering angles using 2D high-$Z_2$ materials such as $MoS_2$. Under these circumstances contributions from Auger transitions following inner-shell excitations in close collisions might be detectable.

**Acknowledgements**

Helpful discussions with other members of the Atomic and Plasma Physics group at TU Wien are gratefully acknowledged. We would also like to thank Barbara Bruckner (Uppsala University) for helping us with the Auger electron spectroscopy measurements. Accelerator operation is supported by the Swedish Research Council VR-RFI (contracts No. 821-2012-5144 and No. 2017-00646_9) and the Swedish Foundation for Strategic Research (contract RIF14-0053).

**References**

[1] G. Fubiani, H.P.L. de Esch, A. Simonin, R.S. Hemsworth, Modeling of secondary emission processes in the negative ion based electrostatic accelerator of the International Thermonuclear Experimental Reactor, Phys. Rev. Spec. Top. - Accel. Beams. 11 (2008) 014202. https://doi.org/10.1103/PhysRevSTAB.11.014202.

[2] G. Kowarik, M. Brunmayr, F. Aumayr, Electron emission from tungsten induced by slow, fusion-relevant ions, Nucl. Instruments Methods Phys. Res. Sect. B. 267 (2009) 2634–2637. https://doi.org/10.1016/j.nimb.2009.05.064.

[3] R. Ramachandra, B. Griffin, D. Joy, A model of secondary electron imaging in the helium ion scanning microscope, Ultramicroscopy. 109 (2009) 748–757. https://doi.org/10.1016/j.ultramic.2009.01.013.

[4] D.C. Joy, B.J. Griffin, Is microanalysis possible in the helium ion microscope?, Microsc. Microanal. 17 (2011) 643–649. https://doi.org/10.1017/S1431927611000596.

[5] G.W. Fraser, The ion detection efficiency of microchannel plates (MCPs), Int. J. Mass Spectrom. 215 (2002) 13–30. https://doi.org/10.1016/S1387-3806(01)00553-X.

[6] E. Gruber, R.A. Wilhelm, R. Pétuya, V. Smejkal, R. Kozubek, A. Hierzenberger, B.C. Bayer, I. Aldazabal, A.K. Kazansky, F. Libisch, A. V. Krasheninnikov, M. Schleberger, S. Facsko, A.G. Borisov, A. Arnau, F. Aumayr, Ultrafast electronic response of graphene to a strong and localized electric field, Nat. Commun. 7 (2016) 13948. https://doi.org/10.1038/ncomms13948.

[7] J. Schwestka, D. Melinc, R. Heller, A. Niggas, L. Leonhartsberger, H. Winter, S. Facsko, F. Aumayr,

R.A. Wilhelm, A versatile ion beam spectrometer for studies of ion interaction with 2D materials, Rev. Sci. Instrum. 89 (2018) 085101. https://doi.org/10.1063/1.5037798.

[8] R.A. Baragiola, E. V. Alonso, A.O. Florio, Electron emission from clean metal surfaces induced by low-energy light ions, Phys. Rev. B. 19 (1979) 121–129. https://doi.org/10.1103/PhysRevB.19.121.

[9] H.D. Hagstrum, Theory of auger ejection of electrons from metals by ions, Phys. Rev. 96 (1954) 336–365. https://doi.org/10.1103/PhysRev.96.336.

[10] F. Aumayr, H. Winter, Potential Electron Emission from Metal and Insulator Surfaces, in: H. Winter, J. Burgdörfer (Eds.), Slow Heavy-Particle Induc. Electron Emiss. from Solid Surfaces, Springer, Berlin, Heidelberg, 2007: pp. 79–112.

[11] H. Kudo, Ion-Induced Electron Emission from Crystalline Solids, Springer, Berlin, Heidelberg, 2001.

[12] H. Eder, M. Vana, F. Aumayr, H.P. Winter, Precise total electron yield measurements for impact of singly or multiply charged ions on clean solid surfaces, Rev. Sci. Instrum. 68 (1997) 165–169. https://doi.org/10.1063/1.1147802.

[13] J. Schwestka, A. Niggas, S. Creutzburg, R. Kozubek, R. Heller, M. Schleberger, R.A. Wilhelm, F. Aumayr, Charge-Exchange-Driven Low-Energy Electron Splash Induced by Heavy Ion Impact on Condensed Matter, J. Phys. Chem. Lett. 10 (2019) 4805–4811. https://doi.org/10.1021/acs.jpclett.9b01774.

[14] H. Rothard, R. Moshammer, J. Ullrich, H. Kollmus, R. Mann, S. Hagmann, T.J.M. Zouros, Differential multi-electron emission induced by swift highly charged gold ions penetrating carbon foils, Nucl. Instruments Methods Phys. Res. Sect. B. 258 (2007) 91–95. https://doi.org/10.1016/j.nimb.2006.12.132.

[15] R. Moshammer, M. Unverzagt, W. Schmitt, J. Ullrich, H. Schmidt-Böcking, A 4 π recoil-ion electron momentum analyzer: A high-resolution "microscope" for the investigation of the dynamics of atomic, molecular and nuclear reactions, Nucl. Instruments Methods Phys. Res. Sect. B. 108 (1996) 425–445. https://doi.org/10.1016/0168-583X(95)01259-1.

[16] M.K. Linnarsson, A. Hallén, J. Åström, D. Primetzhofer, S. Legendre, G. Possnert, New beam line for time-of-flight medium energy ion scattering with large area position sensitive detector, Rev. Sci. Instrum. 83 (2012) 095107. https://doi.org/10.1063/1.4750195.

[17] M.A. Sortica, M.K. Linnarsson, D. Wessman, S. Lohmann, D. Primetzhofer, A versatile time-of-flight medium-energy ion scattering setup using multiple delay-line detectors, Nucl. Instruments Methods Phys. Res. Sect. B. 463 (2020) 16–20. https://doi.org/10.1016/j.nimb.2019.11.019.

[18] S. Lohmann, M.A. Sortica, V. Paneta, D. Primetzhofer, Analysis of photon emission induced by light and heavy ions in time-of-flight medium energy ion scattering, Nucl. Instruments Methods Phys. Res. Sect. B. 417 (2018) 75–80. https://doi.org/10.1016/J.NIMB.2017.08.005.

[19] T. Kobayashi, D. Primetzhofer, M. Linnarsson, A. Hallén, Ion-stimulated desorption in the medium-energy regime, Jpn. J. Appl. Phys. 53 (2014) 060305. https://doi.org/10.7567/JJAP.53.060305.

[20] S. Lohmann, D. Primetzhofer, Ion-induced particle desorption in time-of-flight medium energy ion scattering, Nucl. Instruments Methods Phys. Res. Sect. B. 423 (2018) 22–26. https://doi.org/10.1016/j.nimb.2018.02.016.

[21]  D.J. Manura, D.A. Dahl, SIMION (Version 8.0.3), Scientific Instrument Services Inc. (2007).

[22]  M. Stigsson, Orientation Uncertainty of Structures Measured in Cored Boreholes: Methodology and Case Study of Swedish Crystalline Rock, Rock Mech. Rock Eng. 49 (2016) 4273–4284. https://doi.org/10.1007/s00603-016-1038-5.

[23]  L.E. Davis, N.C. MacDonald, W.C. Palmberg, G.E. Riach, R.E. Weber, Handbook of Auger Electron Spectroscopy, 2nd ed., Physical Electronics Industries, 1976.

[24]  H. Winter, Kinetic Electron Emission for Grazing Scattering of Atoms and Ions from Surfaces, in: H. Winter, J. Burgdörfer (Eds.), Slow Heavy-Particle Induc. Electron Emiss. from Solid Surfaces, Springer, Berlin, Heidelberg, 2007: pp. 113–151.

[25]  S. Tanuma, C.J. Powell, D.R. Penn, Calculations of electron inelastic mean free paths. IX. Data for 41 elemental solids over the 50 eV to 30 keV range, Surf. Interface Anal. 43 (2011) 689–713. https://doi.org/10.1002/sia.3522.

[26]  L. Calliari, S. Fanchenko, M. Filippi, Plasmon features in electron energy loss spectra from carbon materials, Carbon N. Y. 45 (2007) 1410–1418. https://doi.org/10.1016/j.carbon.2007.03.034.

[27]  S. Ritzau, R.A. Baragiola, Electron emission from carbon foils induced by keV ions, Phys. Rev. B. 58 (1998) 2529–2538. https://doi.org/10.1103/PhysRevB.58.2529.

[28]  O. Jagutzki, V. Mergel, K. Ullmann-Pfleger, L. Spielberger, U. Spillmann, R. Dörner, H. Schmidt-Böcking, A broad-application microchannel-plate detector system for advanced particle or photon detection tasks: Large area imaging, precise multi-hit timing information and high detection rate, Nucl. Instruments Methods Phys. Res. Sect. A. 477 (2002) 244–249. https://doi.org/10.1016/S0168-9002(01)01839-3.